\documentclass[a4paper,11pt]{article}

\usepackage{amssymb}
\usepackage{amsmath}
\usepackage{amscd}
\usepackage{latexsym}
\usepackage{epsfig}
\usepackage{color}
\usepackage{fancyhdr}
\usepackage[bf,footnotesize]{caption}
\usepackage{graphicx}
\usepackage{booktabs}
\usepackage{a4wide}
\usepackage{bbm} 
\usepackage{mathrsfs}
\usepackage{dsfont}
\usepackage{wasysym}
\usepackage{array}
\usepackage[all]{xy}
\usepackage{subfigure}

\usepackage{slashed}
\usepackage[numbers,sort&compress]{natbib}
\usepackage[colorlinks=true,a4paper=true,backref=false, linktocpage=true,
citecolor=blue,urlcolor=blue,linkcolor=blue,pdfpagemode=UseOutlines]{hyperref}

\hypersetup{%
  bookmarksnumbered=true,
  pdftitle = {},
  pdfsubject = {},
  pdfauthor = {U.~Wenger},
  pdfkeywords = {}
}

\graphicspath{{Figures_Simulations/}}

\def\min{\mathop{\rm min}}              
\def\sgn{\mathop{\rm sgn}}              
\def\Dslash{\hbox{D}\kern-0.6em\raise0.15ex\hbox{/}} 

\renewcommand{\O}{\mathcal{O}}

\newcommand{\G}{\mathcal{G}}

\newcommand{\Z}{\mathcal{Z}}

\newcommand{\C}{\mathcal{C}}

\newcommand{\psibar}{\overline{\psi}}
\newcommand{\e}{\mathrm{e}}

\def\bi{\begin{itemize}}
\def\ei{\end{itemize}}
\def\be{\begin{equation}}
\def\ee{\end{equation}}

\date{\empty}

\begin{document}
\title{Supersymmetric quantum mechanics on the lattice:\\
III. Simulations and algorithms}

\author{David Baumgartner and Urs Wenger\vspace{0.5cm}
\\
Albert Einstein Center for Fundamental Physics,\\
Institute for Theoretical Physics, University of Bern, \\
Sidlerstrasse 5, CH--3012 Bern, Switzerland\vspace{0.5cm}
\\
}

\maketitle

\begin{abstract}
  In the fermion loop formulation the contributions to the partition
  function naturally separate into topological equivalence classes
  with a definite sign. This separation forms the basis for an
  efficient fermion simulation algorithm using a fluctuating open fermion
  string. It guarantees sufficient tunnelling between the topological
  sectors, and hence provides a solution to the fermion sign problem
  affecting systems with broken supersymmetry. Moreover, the algorithm
  shows no critical slowing down even in the massless limit and can
  hence handle the massless Goldstino mode emerging in the
  supersymmetry broken phase. In this paper -- the third in a series
  of three -- we present the details of the simulation algorithm and
  demonstrate its efficiency by means of a few examples.
\end{abstract}

\section{Introduction}

The reformulation of supersymmetric quantum mechanics on the lattice
in terms of bosonic and fermionic bonds as derived in the first paper
of our series \cite{Baumgartner:2014nka} provides a perfect setup for
Monte Carlo simulations. First of all, the reduction in complexity by
going from continuous to discrete variables is enormous. More
specifically though, expressing the Grassmann fields in terms of
fermionic bonds avoids the expensive calculation of the fermion
determinant and allows the use of special algorithms for which
critical slowing down is essentially absent
\cite{Wenger:2008tq,Wenger:2009mi} and simulations are possible even
in the massless limit \cite{Baumgartner:2011cm}. This is of particular
importance for systems with broken supersymmetry, since the physics of
those is driven by the massless Goldstino mode. In the present paper
-- the last in a series of three -- we describe in detail such an
algorithm and demonstrate its efficiency. Since the model can be
solved exactly at finite lattice spacing by means of transfer
matrices, as discussed in the second paper of our series
\cite{Baumgartner:2015qba}, there is in principle no need for
numerical simulations. Hence, the present paper rather constitutes a
feasibility study to test the practicability and efficiency of the
proposed simulation algorithm for the quantum mechanical system in the
bond formulation. In that sense it also serves as a preparation for
the application of the algorithm, in particular the fermionic part, in
more complex situations, such as in supersymmetric Yang-Mills quantum
mechanics \cite{Steinhauer:2014oda}, in the ${\cal N}=1$ Wess-Zumino
model \cite{Baumgartner:2011jw,Baumgartner:2013ara,Steinhauer:2014yaa}
or in the supersymmetric nonlinear $O(N)$ sigma model
\cite{Steinhauer:2013tba}. The advantage of the application of the
algorithm in the quantum mechanical model presented here is of course
the fact that the correctness of the algorithm can be crosschecked
with the exact results from the transfer matrix approach, and that the
algorithm can hence be validated in detail.

There is another rather pedagogical reason which motivates to consider
a new simulation algorithm for quantum mechanics in the bond
formulation. Often, simple quantum mechanical systems such as the
harmonic and anharmonic oscillator are used to introduce the path
integral approach. Similarly, the systems also provide a pedagogical
context in which various Monte Carlo simulation algorithms can be
illustrated and discussed, see for example \cite{Creutz:1981} for an
early example. However, it turns out that the standard Metropolis
algorithms and even more advanced algorithms such as the
overrelaxation or heat bath algorithm become extremely inefficient
towards the continuum limit. This has to do with the usual critical
slowing down of the simulations towards that limit, and for the
anharmonic oscillator also with the suppressed tunnelling at small
lattice spacing. The algorithms presented here do not suffer from
these deficiencies, because they eliminate critical slowing down.  In
addition, in the bond formulation the $\mathbb{Z}_2$-symmetry $\phi
\rightarrow -\phi$ is exactly maintained for each bond configuration.

Last but not least, the numerical simulations presented here serve as
a test of the practicability of the solution of the fermion sign
problem proposed in \cite{Baumgartner:2011cm} and discussed further in
the first paper of our series \cite{Baumgartner:2014nka}. The solution
is based on two ingredients. Firstly, the lattice regulates the
vanishing Witten index and therefore also the sign problem. Secondly,
the fermion loop formulation provides a tool to handle the fluctuating
sign, because it naturally separates the contributions to the
partition function into topological equivalence classes, each
possessing a definite sign. Nevertheless, it is a priori not clear
whether the lattice artefacts and the statistical fluctuations can be
kept under sufficient control in a practical simulation. The
statistical fluctuations of the sign are essentially determined by the
amount of tunnelling between the topological sectors, i.e., between
the fermionic and bosonic vacuum. In order for the fermion update
algorithm to be a true solution to the sign problem, it must guarantee
a sufficiently efficient tunnelling rate. The results in this paper
demonstrate that this is indeed the case.  Not surprisingly, the open
fermion string algorithm discussed here has proven to be extremely
successful in the ${\cal N}=1$ Wess-Zumino model
\cite{Steinhauer:2014yaa} in which the fermion sign problem is
prevailing.

Of course, supersymmetric quantum mechanics has already been simulated
on the lattice in various setups using standard algorithms, cf.~for
example
\cite{Catterall:2000rv,Giedt:2004vb,Kirchberg:2004vm,Bergner:2007pu,Kastner:2007gz,Wozar:2011gu,Kanamori:2007ye,Kanamori:2007yx,Kanamori:2010gw}. However,
the bond formulation together with the simulation algorithm presented
here brings the numerical nonperturbative calculations to a new,
unprecedented level of accuracy. In that sense, the results presented
here and partly in \cite{Baumgartner:2011cm} serve as a benchmark
against which new formulations or simulation algorithms can be tested.

The present paper is organised as follows. In Section
\ref{sec:algorithm} we construct in detail an algorithm designed for
updating the bosonic and fermionic bond configurations.  The
discussion includes the explicit update steps and the derivation of
the corresponding acceptance ratios. Their evaluation requires the
calculation of site weight ratios which turn out to become numerically
unstable for large site occupation numbers. Therefore, in Section
\ref{sec:weights} we present a computational strategy which allows to
evaluate the ratios for arbitrarily large occupation numbers.  In
Section \ref{sec:results}, we then present the results obtained using
the proposed algorithm. The simulations are for the same
discretisation schemes and superpotentials we used in the previous two
papers \cite{Baumgartner:2014nka,Baumgartner:2015qba}.  Since this
section is merely meant as a validation of the algorithm, the
discussion of the physics behind the results is kept short and we
refer to the exact results in \cite{Baumgartner:2015qba} for a more
thorough discussion.

\section{Simulation algorithm}
\label{sec:algorithm}
We start our discussion from the partition function of supersymmetric
quantum mechanics on the lattice written as a sum over all allowed,
possibly constrained bond configurations $\C = \{n^b_i(x),n^f(x)\}$ in
the configuration space $\Z$,
\begin{equation}
Z = \sum_{\C \subset \Z} W_F(\C) \, ,
\label{eq:Z from bond configurations}
\end{equation}
where the fermion number $F=0,1$ is determined by the fermionic bond
configuration $\{n^f(x)\}$ with $n^f(x)=F, \, \forall x$, and the weight $W_F(\C)$
of a configuration is given by
\begin{equation}
W_F(\C) = \prod_x \left( \prod_i \frac{w_i^{n^b_i(x)}}{n^b_i(x)!} \right)
\prod_x  Q_F(N(x)) \, .
\label{eq:configuration weight}
\end{equation}
Here, $x$ denotes the sites of the lattice and $i$ labels the various
types of bosonic bonds $b_i$ with $i \in \{j \rightarrow k \,| \, j,k
\in \mathbbm{N}\}$. The corresponding bosonic bond weights are denoted
by $w_i$ and $n_i^b(x) \in \mathbbm{N}_0$ is the occupation number of
the bond $b_i$ connecting the sites $x$ and $x+1$. The site weight
$Q_F$ depends on the site occupation number, i.e., the total number of
bosonic bonds connected to site $x$,
\begin{equation}
N(x) = \sum_{j,k} \left(j\cdot n^b_{j \rightarrow k}(x) + k\cdot
  n^b_{j \rightarrow k}(x-1) \right)
\label{eq:site occupation number}
\end{equation}
and is given by
\begin{equation}
\label{eq:site weight}
 Q_F(N) = \int_{-\infty}^\infty d \phi \ \phi^{N} \e^{-V(\phi)}
 M(\phi)^{1 - F} \,.
\end{equation}
In Section \ref{sec:weights} we will discuss in detail the
computational strategy necessary to reliably evaluate ratios of these
integrals for arbitrary and possibly large site occupation
numbers. The type of bonds $b_i$, the weights $w_i$ as well as the
potential $V(\phi)$ and the monomer term $M(\phi)$ in eq.(\ref{eq:site
  weight}) depend on the specifics of the chosen discretisation and
the superpotential $P(\phi)$. We refer to the appendix of our first
paper \cite{Baumgartner:2014nka} for a compilation of the
discretisations and superpotentials considered in our series.

As mentioned above, the bond configurations $\C = \{n^b_i(x),n^f(x)\}$
are possibly constrained. In particular we have the local fermionic
constraints
\begin{equation}
\label{eq:fermionic constraint}
n^f(x-1) = n^f(x) \, 
\end{equation}
while the local bosonic constraints
\begin{equation}
N(x) = 0 \mod 2 \, 
\label{eq:bosonic constraint}
\end{equation}
may or may not be present depending on the bosonic symmetries of the system.

The challenge of updating constrained bond configurations lies
precisely in the difficulty to maintain the constraints while moving
efficiently through the configuration space $\Z$. In
\cite{Prokof'ev:2001} Prokof'ev and Svistunov proposed to extend the
constrained bosonic bond configuration space by introducing local
sources which explicitly violate the constraints. The so-called worm
algorithm then probes the extended configuration space by moving the
local violations around the lattice, thereby sampling directly
the bosonic correlation function corresponding to the sources
introduced. The contact with the original configuration space $\Z$ is
established when the violations annihilate each other, e.g.~when
moving to the same site on the lattice, such that the bond
configuration fulfils again all constraints.

In \cite{Wenger:2008tq} the idea has been extended to fermionic
systems expressed in terms of fermionic bonds. The fermionic
constraint in eq.(\ref{eq:fermionic constraint}) allows only either an
empty or a completely filled fermion bond configuration.  The
difficulty for the direct application of the worm idea to the
fermionic system lies in the fact that the introduction of the fermionic
source term $\psi_x \psibar_x$ is incompatible with the presence of
the fermion loop at site $x$.  A simple solution is to allow the
unphysical situation of the site $x$ being occupied by a propagating
fermion and two additional sources. Such a configuration violates the
Pauli exclusion principle and does not contribute to any physical
observable. In the Grassmann path integral such a configurations
indeed vanishes trivially.

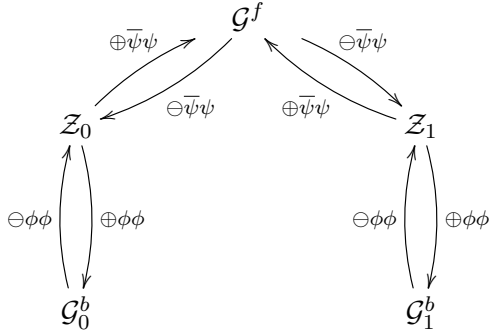
\begin{figure}
\centering
$\xymatrix{
& \qquad \G^f \qquad \ar @/^/ [dl]^{\ominus \psibar \psi}  \ar @/^/[dr]^{ \ominus \psibar \psi}& \\
\Z_0 \ar @/^/[ur]^{\oplus \psibar \psi}  \ar @/^/[dd]^{ \oplus \phi \phi} &  & \Z_1 \ar @/^/ [ul]^{\oplus \psibar \psi}  \ar @/^/[dd]^{\oplus \phi \phi}\\
& & \\
\G^b_0 \ar @/^/[uu]^{\ominus \phi \phi} & & \G_1^b \ar @/^/[uu]^{\ominus \phi \phi} }$
\caption{Schematic representation of the configuration spaces. The
  configuration space $\G^f \equiv \G^f_0 = \G^{\overline{f}}_1$
  mediates between the bosonic and the fermionic sector. By the
  symbols $\oplus$ and $\ominus$, we denote the addition and removal
  of the corresponding source and sink field variables, respectively.}
  \label{fig:conf_diag}
\end{figure}

In order to be more explicit it is necessary to introduce the bond
configuration spaces of bosonic and fermionic two-point correlation
functions, $\G^b_F$ and $\G^f$, respectively, following the notation
in our first paper \cite{Baumgartner:2014nka}. Bond configurations in
$\G^b_F$ contribute to the non-normalised bosonic two-point function
according to
\begin{equation}
\label{eq:g_b}
  g^b_F(x_1 - x_2) 
\equiv \langle \!
  \langle \phi_{x_1} \phi_{x_2} \rangle \! \rangle_F 
  = \sum_{\C \subset
  \G^b_F} \left(\prod_x \frac{Q_F(N(x) +
    \delta_{x,x_1}+\delta_{x,x_2})}{Q_F(N(x))} \right)\cdot W_{F}(\C)
\, ,
\end{equation}
while the configurations in $\G^f$ contribute to the non-normalised
fermionic two-point function as
\begin{equation}
\label{eq:g_f}
  g^f(x_1 - x_2) \equiv \langle
  \! \langle  \psi_{x_1} \psibar_{x_2} \rangle \! \rangle = \sum_{\C \subset \G^f} \left[\prod_{x \in \phantom{\notin} \! \! \! \!
  \mathcal{F}} \frac{Q_1(N(x))}{Q_0(N(x))} \right] \cdot W_0(\C) \, ,
\end{equation}
where $\cal{F}$ denotes the set of lattice sites belonging to the open
fermion string associated with the fermionic correlation function. The
key point of the bosonic and fermionic updating algorithm is that the
bond configurations for $g^b_F(0)$, $g^f(0)$ and $Z_F$ have identical
bond elements.  As a consequence, statistics for $g^{b,f}$ and $Z$ can
be accumulated in the same Monte Carlo process. If the bosonic
constraints in eq.(\ref{eq:bosonic constraint}) are not present,
e.g.~for superpotentials with broken supersymmetry, the equivalence of
bond configurations even extends to $g^b_F(x)$, i.e., $\Z_F =
\G^b_F$. The movements from one configuration space to the other are
induced by introducing or removing bosonic or fermionic sources
according to the scheme given in figure 8 of our first paper
\cite{Baumgartner:2014nka}. For convenience we reproduce it here in
figure \ref{fig:conf_diag}.

In the following we will now discuss in detail the various updating
steps which establish explicitly the connection between the bond
configuration spaces $\G^f, \Z_F, \G^b_F$ and in addition move the
system within $\G^f$ and $\G^b_F$. The moves are generated by a Monte
Carlo process with probabilities given by the weights of the
configurations in eqs.(\ref{eq:configuration weight}), (\ref{eq:g_b})
and (\ref{eq:g_f}). In particular, we derive the transition
probabilities $P_X(\C \rightarrow \C^\prime)$ for the transition $X$
from bond configuration $\C$ to $\C'$, which is then accepted by the
usual Metropolis prescription
\begin{equation}
 P_{acc}(\C \rightarrow \C^\prime) = \min\{1,P_X(\C \rightarrow \C^\prime)\}.
\end{equation}
In order to simplify the discussion we select the update from $\Z_F$
to $\G^b_F$ or $\G^f$ with equal probability which is balanced by
corresponding proposal probabilities to select between moving in
$\G^f$ and $\G^b_F$ or returning to $\Z_F$.

\subsection{Updating the fermionic bond configuration}
Here we discuss the various update steps which moves the system within
$\G^f$ and relate the bond configurations spaces $\Z_0$ and $\Z_1$ via
$\G^f$.

\begin{figure}[h]
\centering
\includegraphics[width = 0.65\textwidth]{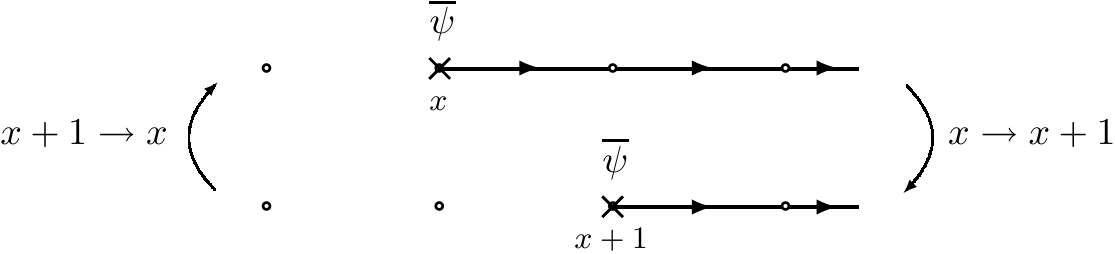}
 \caption{Fermionic bond configuration update algorithm. Graphical
   representation of the `shift' update step $x \rightarrow x+1$ in
   forward direction for an open fermion string configuration. It
   is balanced with the shift update step $x+1 \rightarrow x$ in
   backward direction. The bosonic background bond configuration is not drawn.}
  \label{fig:fer_update_shift}
\end{figure}

Moves within $\G^f$ are induced by shifting $\psibar$ by one lattice
spacing from site $x$ to site $x+1$, and vice versa, while keeping the
other source $\psi$ fixed. Such an update step is graphically
illustrated in figure \ref{fig:fer_update_shift} and is called `shift'
update step.
A shift in forward direction, $x \rightarrow x + 1$, automatically
involves the removal of the fermionic bond $b^f(x)$, whereas a shift
in backward direction, $x+1 \rightarrow x$, requires the addition of a
new fermionic bond $b^f(x)$. Both directions are proposed with equal
probability $1/2$ and are hence balanced against each other as long as
the new site does not coincide with the position of the source
$\psi$. The formula in eq.(\ref{eq:g_f}) provides us with the
acceptance ratios
\begin{eqnarray}\label{acc_shift_f}
 P_{\mathrm{sh}}(x + 1 \rightarrow x)& = &
\displaystyle \frac{Q_1(N(x))}{Q_0(N(x))}\, ,\\[6pt]
 P_{\mathrm{sh}}(x \rightarrow x+1 )& = & \displaystyle
 \frac{Q_0(N(x))}{Q_1(N(x))} \,.
\label{acc_shift_b}
\end{eqnarray}

Now let us consider the case when we propose to shift the source
$\psibar$ forward to the site $x$ where the source $\psi$ is present,
as depicted in the upper half of figure \ref{fig:fermionic put_hybrid
  step}.
\begin{figure}[h]
\centering
\includegraphics[width = 0.6\textwidth]{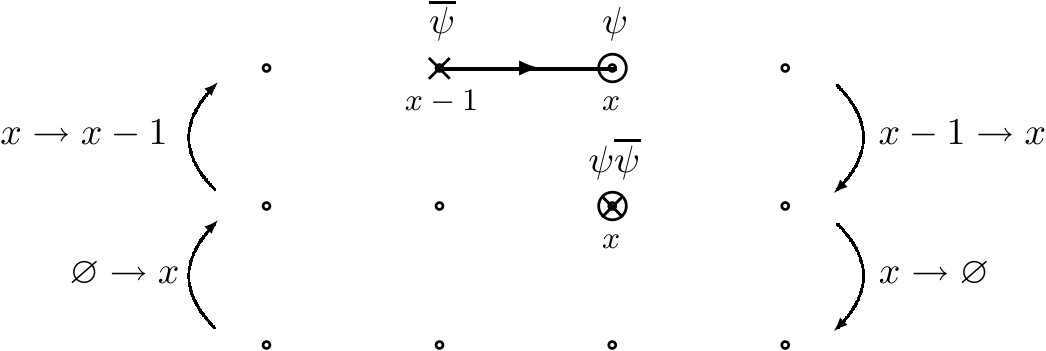}
\caption{Fermionic bond configuration update algorithm. Graphical
  representation of the `shift' update step $x \rightarrow x-1$ or
  $x-1 \rightarrow x$, respectively, and the `put/remove' update step
  $\varnothing \rightarrow x$ and $\varnothing \rightarrow x$. The
  sources are marked with a $\ocircle $ for $\psi$ and a $\times$ for
  $\psibar$. The bosonic background bond configuration is not drawn.}
  \label{fig:fermionic put_hybrid step}
\end{figure}

The forward shift update step $x-1 \rightarrow x$ is balanced with the
backward shift update step $x \rightarrow x-1$. This backward shift,
however, is proposed with probability 1 instead of probability 1/2
since the shift of $\psibar$ from $x \rightarrow x+1$ would involve
the creation of an open fermion string around the entire lattice. The
asymmetry in the proposition probabilities is balanced by the choice
of the probability $p_{\mathrm{rm}} = 1/2$ to remove the sources
$\psi \psibar$, such that we find the acceptance ratio for a shift in
backward direction $x \rightarrow x-1$ to be the same as in
eq.(\ref{acc_shift_f}), namely
\begin{equation}
P_{\mathrm{sh}}(x \rightarrow x-1) =  \frac{Q_1(N(x-1))}{Q_0(N(x-1))}.
\end{equation}
The shift step is balanced with the corresponding one in forward
direction with acceptance ratio as given in eq.(\ref{acc_shift_b}),
\begin{equation}
 P_{\mathrm{sh}}(x-1 \rightarrow x) = 
\displaystyle \frac{Q_0(N(x-1))}{Q_1(N(x-1))} 
\end{equation}

The step from $\G^f$ to $\Z_0$ and vice versa is induced by
introducing or removing a pair of fermionic sources $\psi\psibar$
at site $x$, respectively. It is called `put/remove' update step and
is graphically illustrated in the lower half of figure
\ref{fig:fermionic put_hybrid step}.
The removal of the fermionic sources is suggested with probability
$p_{\mathrm{rm}} = 1/2$ and is balanced on one side by the probability
to add bosonic sources, and on the other by the probability to shift
one of the sources and hence move within $\C^f$.  Because the
`put/remove' update step does not alter the fermionic bond
configuration, we have $\Z_0 = \G^f(0)$. On the other hand it adds or
removes a fermionic monomer term $M(\phi)$ at site $x$. The relative
weight of the configurations with or without this term is given by
$Q_1/Q_0$ and the acceptance ratios on a lattice with $L_t$ sites
become
\begin{eqnarray}
 P_{\mathrm{rm}}(x \rightarrow \varnothing) & = &
 \frac{2}{L_t}\displaystyle \frac{Q_0(N(x))}{Q_1(N(x))}, \\[6pt]
 P_{\mathrm{put}}(\varnothing \rightarrow x) & = &
 \frac{L_t}{2}\displaystyle \frac{Q_1(N(x))}{Q_0(N(x))}\,.
\end{eqnarray}
The factor $L_t$ compensates for the proposition probability to choose
lattice site $x$ out of $L_t$ possibilities when putting the sources,
while the factor 2 compensates for the asymmetric shift proposal
probability when moving $\psibar$ from $x$ to $x-1$, since the
shift of $\psibar$ from $x$ to $x+1$ is not allowed.

\begin{figure}[h]
\centering
\includegraphics[width = 0.65\textwidth]{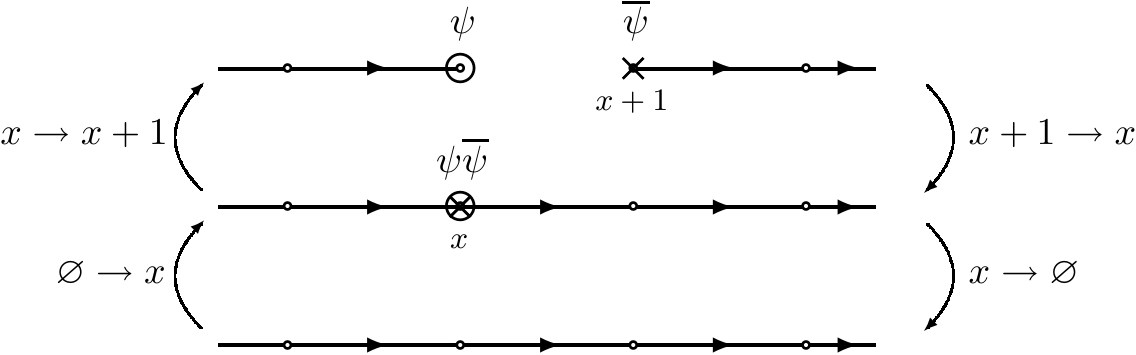}
\caption{Fermionic bond configuration update algorithm. Graphical
  representation of the hybrid `shift/remove' update step $x+1
  \rightarrow x \rightarrow \varnothing$ in backward direction,
  balanced with the `put/shift' update step, $\varnothing \rightarrow
  x \rightarrow x+1$. The bosonic background bond configuration is not
  drawn.}
  \label{fig:fer_update_shift_hybrid}
\end{figure}

Next we consider the shift update step for the case when the source
$\psibar$ at site $x+1$ is shifted backwards to site $x$ which is
already occupied by the sink $\psi$. The step is graphically
illustrated in the upper half of figure
\ref{fig:fer_update_shift_hybrid}.
While the resulting fermion bond configuration is a valid one (it
belongs to $\Z_1$), the whole fermion configuration including the
source and the sink represents an unphysical situation, and in fact
does not contribute to any physical observable, as discussed
before. Therefore, such a backward shift from $\G^f$ to $\Z_1$,
essentially closing the open fermion string, automatically induces the
removal of the fermionic source and sink pair $\psi\psibar$ from site
$x$ as illustrated in the lower half of figure
\ref{fig:fer_update_shift_hybrid}. Such a step is called a hybrid
`shift/remove' update step. Of course, the step is balanced with a
hybrid `put/shift' update step when the additional fermionic sink and
source variables are put on a closed fermion loop at the site $x$. As
usual, the acceptance ratios for the hybrid update steps can be read
off from the weights of the configurations involved and yield
\begin{eqnarray}
  P_{\mathrm{sh/rm}}(x+1 \rightarrow x \rightarrow \varnothing) & = & \frac{2}{L_t}, \\[6pt]
  P_{\mathrm{put/sh}}(\varnothing \rightarrow x \rightarrow x+1) & = & \frac{L_t}{2}.
\end{eqnarray}
The factor $L_t$ compensates for the proposition probability to choose
the same lattice site $x$ when putting the sources $\psi\psibar$ back 
on the lattice, whereas the factor $2$ compensates for the proposition
probability to shift in forward or backward direction when the fermion
string is still open. Note that there are no ratios of $Q$-weights
involved, since no monomer term is added or removed by the hybrid
shift/remove update step.

To complete our discussion of the fermionic bond update, we note that
the algorithm provides improved estimators for the fermionic two-point
function $g^f(x)$ and the partition functions $Z_F$. Because the
algorithm samples directly the configuration space $\G^f$, every open
fermion string configuration contributes unity to the stochastic Monte
Carlo estimator for $g^f(x)$. To be precise we have
\begin{equation}
\label{eq:g^f estimator}
 g^f(x|\C \in \G^f) = \delta_{x,x_1 - x_2} \,,
\end{equation} 
where $x_1$ and $x_2$ are the end and starting point of the open
fermion string, i.e., the positions of the sink $\psi$ and the source
$\psibar$, respectively.  Similarly, every bond configuration in
$\Z_F$ is generated with its proper weight and hence contributes unity
to the stochastic estimator for $Z_F$, i.e., the Monte Carlo
estimator for $Z_F$ is simply
\begin{equation}
Z_F(\C \in \Z_F) = 1\,.
\end{equation}

Finally we note that the factors of $L_t$ appearing in the acceptance
ratios above may become inconvenient in practice, especially towards
the continuum limit when $L_t\rightarrow \infty$. The factors only
occur when contact between $\Z_F$ and $\G^f$ is made, i.e., they are
responsible for getting the relative normalisation between $Z_F$ and
$g^f$ right. However, since we make use of translational invariance in
eq.(\ref{eq:g^f estimator}) the factors of $L_t$ are in fact cancelled
and can hence be omitted.

\subsection{Updating the bosonic bond configuration}
In this section we now discuss the update steps which relate the bond
configuration spaces $\Z_F$ and $\G^b_F$ for a fixed fermionic bond
configuration with fermion number $F=0,1$.

We point out that for an arbitrary superpotential there are in general
no restrictions on the bosonic bond configurations. This is for
example the case for the superpotential $P_b$ which we consider in our
series of papers, cf.~eq.(\ref{eq:superpotential P_b}). In contrast,
the superpotential $P_u$ in eq.(\ref{eq:superpotential P_u}) yields
the local constraint $N(x) = 0 \mod 2$ on the site occupation number,
due to the parity symmetry $\phi \rightarrow -\phi$. In the following
discussion, we always present the generic case first, and then specify
the modifications or simplifications due to the constraint.  In
analogy to the fermionic bond update, the `put/remove' and the `shift'
updates are the main steps for updating the bosonic bond configurations.
The `put/remove' step introduces or removes one or two sources $\phi$,
while the `shift' step shifts the sources by one lattice spacing.  If
there are no restrictions on the bond configuration, we are free to
decide for each Monte Carlo step whether to proceed by a `remove' update
or a `shift' update.  With probability $p_{\mathrm{rm}}$, we propose to
remove the sources from the lattice, while the proposition to continue
the worm update with a `shift' step is chosen with probability $1 -
p_{\mathrm{rm}}$.

The step from $\G_F^b$ to $\Z_F$ (and vica versa) is induced by
removing (or introducing) a bosonic source $\phi$ at sites $x_1$ and
$x_2$, with $x_1=x_2$ not excluded.  The step does not alter the bond
configuration, but only the site occupation numbers at sites $x_1$ and
$x_2$. Thus, only the ratios of the site weights $Q_F$ are involved in
the acceptance probability for e.g.~the `remove' step,
\begin{eqnarray}\label{eq_remove_boson}
 P_{\mathrm{rm}}(x_1,x_2 \rightarrow \varnothing)\!\!\! & = & \!\!\!\left\{ 
\begin{array}{ll}
\displaystyle  \frac{1}{p_{\mathrm{rm}}L_t^2} \frac{Q_F(N(x_1)-2)}{Q_F(N(x_1))} & \mathrm{if} \ x_1 = x_2,\\[20pt]
\displaystyle  \frac{1}{p_{\mathrm{rm}}L_t^2} \frac{Q_F(N(x_1)-1)}{Q_F(N(x_1))} \frac{Q_F(N(x_2)-1)}{Q_F(N(x_2) )} & \mathrm{if} \ x_1 \neq x_2.
\end{array} \right. 
\end{eqnarray}
The prefactor $1/(p_{\mathrm{rm}}L_t^2)$ is motivated as follows. The
factor $1/L_t^2$ balances the probability for the proposition of putting
the bosonic sources at the sites $x_1$ and $x_2$ when re-entering the
configuration space $\G^b_F$, while the factor $1/p_{\mathrm{rm}}$
balances the proposition probability for the choice of proceeding by
the shift update instead of the remove update, as discussed above.
The acceptance ratios for re-entering the configuration space $\G^b_F$
from $\Z_F$ are given by
\begin{eqnarray}
 P_{\mathrm{put}}(\varnothing \rightarrow x_1,x_2) & = &\left\{ \begin{array}{ll}
\displaystyle p_{\mathrm{rm}}L_t^2 \frac{Q_F(N(x_1) + 2)}{Q_F(N(x_1))} & \mathrm{if} \ x_1 = x_2,\\[20pt]
\displaystyle p_{\mathrm{rm}}L_t^2 \frac{Q_F(N(x_1) + 1)}{Q_F(N(x_1))} \frac{Q_F(N(x_2) + 1)}{Q_F(N(x_2))}& \mathrm{if} \ x_1 \neq x_2.
\end{array} \right. \qquad
\label{eq:bosonic remove update}
\end{eqnarray}

Two remarks are in order. Firstly, if there are no constraints on the
bond configuration, one can in principle introduce just a single
source $\phi$ which subsequently is shifted around. In effect, the
algorithm then samples the one-point function which in this situation
is indeed nonvanishing. Secondly, we note that if the constraint $N =
0 \mod 2$ is in place, the two sources can only be placed or removed
when $x_1 = x_2$. As a consequence, only the first of the two
acceptance ratios in eq.(\ref{eq_remove_boson}) and
eq.(\ref{eq:bosonic remove update}) are relevant, while the second
ones are zero by definition.

Next, we discuss the bosonic `shift' update.  With this step we now
change the bosonic bond configuration.  Shifting the source from site
$x$ to a next neighbouring site $y$ is always associated with an
increase or a decrease of the bosonic bond occupation number between
the sites $x$ and $y$ by one. Whether or not the occupation number is
increased or decreased is decided with probability $1/2$. Similarly,
the source can move forward or backward, and we propose both
directions with equal probability $1/2$.  In addition, when there are
several types of bosonic bonds $b_i$ with \mbox{$i \in \{j\rightarrow
  k | j,k \in \mathbb{N}\}$}, we need to decide in each step which
bond is updated. We do so by choosing the proposition probabilities
$p_{j\rightarrow k}$ with $\sum_{j,k} p_{j\rightarrow k} =
1$. However, because the proposals are completely symmetric, these
probabilities do not affect the acceptance ratios.  In the following,
we will use the shorthand notation
\begin{eqnarray}\label{abbrev_n_11}
 n^{j\rightarrow k}_{xy} & = &\left\{ \begin{array}{ll}
n^b_{j \rightarrow k}(x) & \mathrm{if} \ y = x + 1,\\
n^b_{j \rightarrow k}(y)& \mathrm{if} \ y = x - 1,
\end{array} \right. \qquad
\end{eqnarray}
for the occupation number of the bosonic bonds $b_{j\rightarrow k}$
between the sites $x$ and $y$. The shifts $x \rightarrow y$ and
$n^{j\rightarrow k}_{xy} \rightarrow n^{j\rightarrow k}_{xy} + 1$ are
balanced with shifts $y \rightarrow x$ and $n^{j\rightarrow k}_{xy}
\rightarrow n^{j\rightarrow k}_{xy} - 1$, which gives the acceptance
ratios
\begin{multline}
  P_{\mathrm{sh}}(x \rightarrow y, n^{j\rightarrow k}_{xy} \rightarrow
  n^{j\rightarrow k}_{xy} + 1) \!  \\
 = \!
  \left\{\!\!\! \begin{array}{ll} 
 \displaystyle  \frac{w_{j\rightarrow k}}{n^{j\rightarrow k}_{xy} + 1}
\frac{Q_F(N(x)+j-1)}{Q_F(N(x))} \cdot \frac{Q_F(N(y)+k+1)}{Q_F(N(y))} 
 & \mathrm{if} \ y = x + 1,\\[12pt]
\displaystyle    \frac{w_{j\rightarrow k}}{n^{j\rightarrow k}_{xy} +
  1} \frac{Q_F(N(x)+k-1)}{Q_F(N(x))} \cdot \frac{Q_F(N(y)+j+1)}{Q_F(N(y))}
 & \mathrm{if} \ y = x - 1,\\
 \end{array} \right.
\end{multline}
\begin{multline}
  P_{\mathrm{sh}}(x \rightarrow y, n^{j\rightarrow k}_{xy} \rightarrow
  n^{j\rightarrow k}_{xy} - 1) \\
  = \! \left\{ \!\!\! \begin{array}{ll}
\displaystyle   \frac{n^{j\rightarrow k}_{xy}}{w_{j\rightarrow k}}
\frac{Q_F(N(x)-j-1)}{Q_F(N(x))} \cdot \frac{Q_F(N(y)-k+1)}{Q_F(N(y))} 
& \mathrm{if} \ y = x + 1,\\[12pt]
\displaystyle    \frac{n^{j\rightarrow k}_{xy}}{w_{j\rightarrow k}}
\frac{Q_F(N(x)-k-1)}{Q_F(N(x))} \cdot \frac{Q_F(N(y)-j+1)}{Q_F(N(y))}
 & \mathrm{if} \ y = x - 1.
 \end{array} \right.
\end{multline}
Of course, these generic ratios simplify considerably for the specific
bonds $b_i, i \in \{1\rightarrow 1, 1 \rightarrow 2, 1 \rightarrow
3\}$ relevant for the superpotentials considered in our series of
papers.  For example, the acceptance ratios for updating the bond
$b_{1 \rightarrow 1}$ read
\begin{eqnarray}\label{p_sh_bos_11_1}
  P_{\mathrm{sh}}(x \rightarrow y, n^{1\rightarrow 1}_{xy} \rightarrow n^{1\rightarrow 1}_{xy} + 1)
  & = & \frac{w_{1\rightarrow 1}}{n^{1\rightarrow 1}_{xy} + 1}\cdot \frac{Q_F(N(y)+2)}{Q_F(N(y))}, \\ \label{p_sh_bos_11_2}
  P_{\mathrm{sh}}(x \rightarrow y, n^{1\rightarrow 1}_{xy} \rightarrow n^{1\rightarrow 1}_{xy} - 1)
  & = & \frac{n^{1\rightarrow 1}_{xy}}{w_{1\rightarrow 1}} \cdot \frac{Q_F(N(x) - 2)}{Q_F(N(x))} \, .
\end{eqnarray}
Because the bond is symmetric, there is no need to distinguish whether
$y=x+1$ or $y=x-1$.

To complete the discussion of the bosonic bond update, we point out
that the algorithm again provides improved estimators for the bosonic
two-point function $g^b_F(x)$ and the partition functions $Z_F$. As in
the fermionic case, the algorithm samples directly the configuration
space $\G^b_F$ with the correct weighting when the sources are
present. Therefore, every configuration contributes unity to the
stochastic Monte Carlo estimator for $g^b_F(x)$, and we have
\begin{equation}
g_F^b(x| \G^b_F) = \delta_{x_1 - x_2,x} \, ,
\end{equation}
where $x_1$ and $x_2$ are the positions of the two sources. Whenever
the bosonic update decides to remove the sources, we have a
configuration in $\Z_F$ and hence a contribution of unity to the
stochastic estimator for $Z_F$, that is, we have
\begin{equation}
Z_F(\C \in \Z_F) = 1\,.
\end{equation}

In complete analogy to the fermionic update we note that the factors
of $L_t$ appearing in the acceptance ratios of the `put/remove' step
can be compensated by adjusting the overall normalisation of the
two-point function, e.g.~by making use of translational invariance.

\section{Calculation of the site weight ratios}\label{sec:weights}
In order to calculate the weight of a bond configuration, it is
necessary to know the site weights
\begin{equation}
 Q_F(n)  = \int_{-\infty}^\infty d \phi \  \phi^n \e^{-V(\phi)}M(\phi)^{1 - F},
\end{equation}
where $V(\phi)$ and $M(\phi)$ depend on the superpotential and the
discretisation employed, and $F = 0,1$ is the fermion number, for
arbitrary values of the site occupation number $n$. The values of $n$
required in practice are usually limited to $\O(10^3)$.  However, it
turns out that even for moderate values of $n$ of order $\O(100)$ the
site weights $Q_F(n)$ can quickly grow larger than $10^{100}$ or
more. As a consequence, the calculation of the site weights quickly
becomes numerically unstable for growing $n$. In fact, even for simple
potentials when the weights can be calculated analytically in terms of
confluent hypergeometric functions, the numerical evaluation of these
functions is difficult for large $n$, and even specialised libraries
such as the ones available in Wolfram's Mathematica
\cite{mathematica8.0} appear not to be accurate enough.

Fortunately, for the Monte Carlo simulations we only need ratios of
the site weights, such as $Q_F(n+1)/Q_F(n), Q_F(n+2)/Q_F(n)$ and
$Q_1(n)/Q_0(n)$, and these ratios usually do not become larger than
$\O(10)$ even for large $n$. In addition, also the transfer matrix
elements can be rewritten in terms of these ratios as discussed in the
appendix of our second paper of the series \cite{Baumgartner:2015qba}.
Therefore, we now present a numerically stable computational strategy
to calculate the site weight ratios reliably for arbitrary values of
the site occupation numbers.

We start by defining an arbitrary polynomial superpotential
\begin{equation}
 P(\phi) = \sum_{i=0}^p p_i\phi^i,
\end{equation}
and the corresponding bosonic self-interaction potential $V(\phi)$ as
well as the monomer weight $M(\phi)$,
\begin{equation}
  V(\phi)  =  \sum_{i=0}^{2(p-1)} k_i\phi^i, \quad\quad
  M(\phi)  =  \sum_{i=0}^{p-2} m_i\phi^i. 
\end{equation}
Explicitly, the weights in each sector are then given by
\begin{equation}\label{moments}
 Q_1(n) =  \int_{-\infty}^\infty d \phi \ \phi^n \e^{-V(\phi)}
\end{equation}
and
\begin{equation}
 Q_0(n) =  \sum_{i=0}^{p-2}m_iQ_1(n + i).
\end{equation}
For convenience we also define the ratios of the site weights
$Q_F(n)$,
\begin{eqnarray}
  R'_F(n) & = & \frac{Q_F(n + 1)}{Q_F(n)}, \\
  R_F(n) & = & \frac{Q_F(n + 2)}{Q_F(n)}, \\ \label{R_pmf}
 R_m(n) & = & \frac{Q_0(n)}{Q_1(n)}
\end{eqnarray}
which are used for the acceptance ratios in the Monte Carlo
simulations. In principle, only the ratios $R_1'(n)$ need to be
calculated since all other ratios can be derived from those. For
example, $R_1(n)$ can be expressed in terms of $R_1'(n)$ as
\begin{equation}\label{R0}
 R_1(n) = R_1'(n + 1)R_1'(n) \, , 
\end{equation}
but since in some cases $Q_1(n \, \text{odd}) = 0$ the introduction of
$R_1(2n)$ is nevertheless necessary. $R_m(n)$ can be expressed via the
ratios $R_1(n)$ and $R_1'(n)$ and appropriate products thereof,
\begin{equation}\label{R_m}
 R_m(n) = m_0 + R_1'(n) \left(m_1 + R_1(n+2) \left(m_3+\ldots\right)\right) +  R_1(n) \left(m_2
 + R_1(n+2) \left(m_4 + \ldots\right)\right) ,
\end{equation}
and the ratios $R_0'(n)$ and $R_0(n)$ via $R_m(n)$, $R_1(n)$ and $R_1'(n)$ by
\begin{eqnarray}\label{R_1}
 R_0(n) & = & \frac{R_m(n + 2)}{R_m(n)}R_1(n), \\ \label{R_1p}
 R_0'(n) & = & \frac{R_m(n + 1)}{R_m(n)}R_1'(n).
\end{eqnarray}

First, we now discuss how to gain numerical stability for the special
case of an even superpotential $P(\phi)$. In a second step we will
then adapt the idea to treat the somewhat more subtle case of an
arbitrary superpotential.

\subsection{Even superpotential}
Unbroken supersymmetric quantum mechanics requires a superpotential
$P(\phi)$ with $\deg(P(\phi))$ = 0 mod 2. In particular, in our series
of papers we investigate the superpotential
\begin{equation}
P(\phi) = p_2\phi^2 + p_4\phi^4
\end{equation}
which is symmetric w.r.t.~the parity transformation $\phi \rightarrow
-\phi$. As a consequence of the symmetry, $Q_F(n \, \text{odd}) = 0$
for both $F=0,1$ and the ratios $R_F'(n)$ need not be considered --
instead, it is sufficient to determine $R_1(2n)$ with $n \in
\mathbb{N}_0$ only.

For the the potential $V(\phi)$ we then have the form
\begin{equation}
 V(\phi) = k_2\phi^2 + k_4\phi^4 + k_6\phi^6\, ,
\end{equation}
consistent with both the standard discretisation and the $Q$-exact
one.  To keep the integrals numerically under control, for fixed $n$
we apply a variable transformation $\phi \rightarrow \phi /
\widetilde{\phi}$ to obtain rescaled weights $\widetilde{Q}_1(2n)$ as
\begin{equation}
 Q_1(2n) = \widetilde{\phi}\,^{2n + 1}\widetilde{Q}_1(2n).
\end{equation}
Since we have $Q_1(2n) \geq 0$, we can choose the rescaling factor to
be $\widetilde{\phi} = Q_1(2n)^{1/(2n + 1)}$ and the rescaled weight
becomes $\widetilde{Q}_1(2n) = 1$. Calculating the ratio of rescaled
weights as
\begin{equation}
 \widetilde{R}_1(2n) = \frac{\widetilde{Q}_1(2n + 2)}{\widetilde{Q}_1(2n)} = \widetilde{Q}_1(2n + 2),
\end{equation}
where both integrals $\widetilde{Q}_1(2n + 2)$ and
$\widetilde{Q}_1(2n)$ are rescaled with the same factor
$\widetilde{\phi} = Q_1(2n)^{1/(2n + 1)}$, we find that
\begin{equation}
 R_1(2n) = \widetilde{\phi}\,^2 \, \widetilde{R}_1(2n) \, .
\end{equation}
In addition, the rescaled weight $\widetilde{Q}_1(2n + 2)$ is now of
$\O(1)$ and can be evaluated reliably via numerical integration.  So
if we start by integrating directly the numerically stable site
weights $Q_1(0)$ and $Q_1(2)$, we can recursively generate ratios
$R_1(2n)$ with higher and higher $n$. Note that after each calculation
of a ratio $R_1(2n)$, one needs to update the rescaling factor
$\widetilde{\phi} \rightarrow \widetilde{\phi}\,'$. This can be
achieved most easily via
\begin{equation}
 \widetilde{\phi}\,' = \widetilde{\phi}\,^{\frac{2n + 1}{2n + 3}}\,R_1(2n)^{\frac{1}{2n + 3}}.
\end{equation}
Our procedure guarantees that all involved quantities are of
$\O(1)$. Once all ratios $R_1(2n)$ are known, one can calculate the
ratios $R_m(2n)$, noting that for the specific superpotential we
consider, eq.(\ref{R_m}) simplifies to
\begin{equation}
 R_m(2n) = m_0 + m_2R_1(2n).
\end{equation}
The calculation of the ratios $R_0(2n)$ as given in eq.(\ref{R_1}) is
then straightforward.

\subsection{Arbitrary Superpotential}
\label{sec:weights arbitrary superpotential}
In the context of broken supersymmetric quantum mechanics, one
encounters superpotentials with $\deg(P(\phi))$ = 1 mod 2. Therefore,
we now adapt the procedure from above to superpotentials of this
form. For simplicity, we restrict ourselves to the odd superpotential
we consider as the example in our series of papers,
\begin{equation}
 P(\phi) = \sum_{i = 1}^3 p_i\phi^i.
\end{equation}
If at least one of the coefficients $p_1$ and $p_2$ is nonzero, which
is always the case for the superpotentials we use, $V(\phi)$ reads
\begin{equation}
 V(\phi) = k_1\phi + k_2\phi^2 + k_3\phi^3 + k_4\phi^4,
\end{equation}
and at least one of the coefficients $k_1$ and $k_3$ is nonzero
either. This has a two important consequences. Firstly, the moments
defined in eq.(\ref{moments}) are nonzero for $n$ odd, from which it
follows that the ratios $R_F'(n)$ defined in eq.(\ref{R_pmf}) have to
be calculated as well. Secondly, the weights $Q_1(n)$ are no longer
necessarily positive. It turns out, however, that for all practical
purposes it does not affect the simulations. We will discuss this
further in Section \ref{sec:results}.

For the evaluation of the integrals, we apply the same variable
transformation $\phi \rightarrow \phi / \widetilde{\phi}$ as before,
such that we have rescaled weights $\widetilde{Q}_1(n)$ given by
\begin{equation}
 Q_1(n) = \widetilde{\phi}\,^{n + 1}\,\widetilde{Q}_1(n).
\end{equation}
We now choose $\widetilde{\phi} = |Q_1(n)|^{1/(n + 1)} \cdot
\sgn(Q_1(n))$. Then, the integral becomes $\widetilde{Q}_1(n) = 1$
again as before. Furthermore, defining the rescaled ratios
$\widetilde{R}_1'(n)$ to be
\begin{equation}
 \widetilde{R}_1'(n) = \frac{\widetilde{Q}_1(n + 1)}{\widetilde{Q}_1(n)} = \widetilde{Q}_1(n + 1),
\end{equation}
where both integrals $\widetilde{Q}_1(n + 1)$ and $\widetilde{Q}_1(n)$
are rescaled with the same factor $\widetilde{\phi} = |Q_1(n)|^{1/(n +
  1)}\cdot \sgn(Q_1(n))$, we find $R_1'(n) = \widetilde{\phi}
\,\widetilde{R}_1'(n)$. We proceed analogously to the case of the even
superpotential by recursive iteration, with the only exception that we
generate the ratios $R_1'(n)$ instead of the ratios $R_1(n)$. The
update for the rescaling factor $\widetilde{\phi} \rightarrow
\widetilde{\phi}\,'$ is done via
\begin{equation}
 \widetilde{\phi}\,' = |\widetilde{\phi}|^{\frac{n + 1}{n + 2}}\, |R_1'(n)|^{\frac{1}{n + 2}} \cdot \sgn(R_1'(n)).
\end{equation}
Once all the ratios $R_1'(n)$ are known, one can calculate the ratios
$R_1(n)$ via eq.(\ref{R0}), the ratios $R_m(n)$ via eq.(\ref{R_m}),
and the ratios $R_0(n)$ and $R_0'(n)$ via eq.(\ref{R_1}) and
(\ref{R_1p}), respectively.

\section{Results of the Monte Carlo simulations}
\label{sec:results}
The results in this section are merely thought of as a proof of the
feasibility of the algorithm and as a test of its efficiency.
Comparing the Monte Carlo results with the exact solution of the
system at finite lattice spacing provided in our second paper
\cite{Baumgartner:2015qba} of course also serves as a validation for
the algorithm. We refer to that paper for a thorough discussion and
physical interpretation of the results.

For the following Monte Carlo simulations, we consider the same
superpotentials and discretisations as in the previous two papers. In
particular, we simulate the system using the action with counterterm
for both unbroken and broken supersymmetry as well as the $Q$-exact
action for unbroken supersymmetry. Details for the various actions can
be found in the first paper of our series.  Here we only give the
details of the superpotentials for broken and unbroken supersymmetry,
respectively,
\begin{align}
\label{eq:superpotential P_b}
P_b(\phi) &= -\frac{\mu^2}{4 \lambda}\phi + \frac{1}{3} \lambda \phi^3 \, , \\
 P_u(\phi) &= \frac{1}{2}\mu \phi^2 + \frac{1}{4}g \phi^4 \, ,
\label{eq:superpotential P_u}
\end{align}
and we recall that the continuum limit is taken by fixing the
dimensionful parameters $\mu, g, \lambda$ and $L$ while taking the
lattice spacing $a \rightarrow 0$. In practice, the dimensionless
ratios $f_u = g/\mu^2$ and $f_b = \lambda/\mu^{3/2}$ fix the couplings
and $\mu L$ the extent of the system in units of $\mu$, while $a \mu$
and $a/L$ are subsequently sent to zero.  In analogy to the number of
sweeps for a standard Monte Carlo simulation, we count the number of
times the algorithm is in either one of the two configuration spaces
$\Z_F, F=0,1$. The statistics for a simulation is therefore given by
$Z_0 + Z_1 = Z_{a}$.

\begin{figure}
 \centering
 \includegraphics[width = 0.8\textwidth]{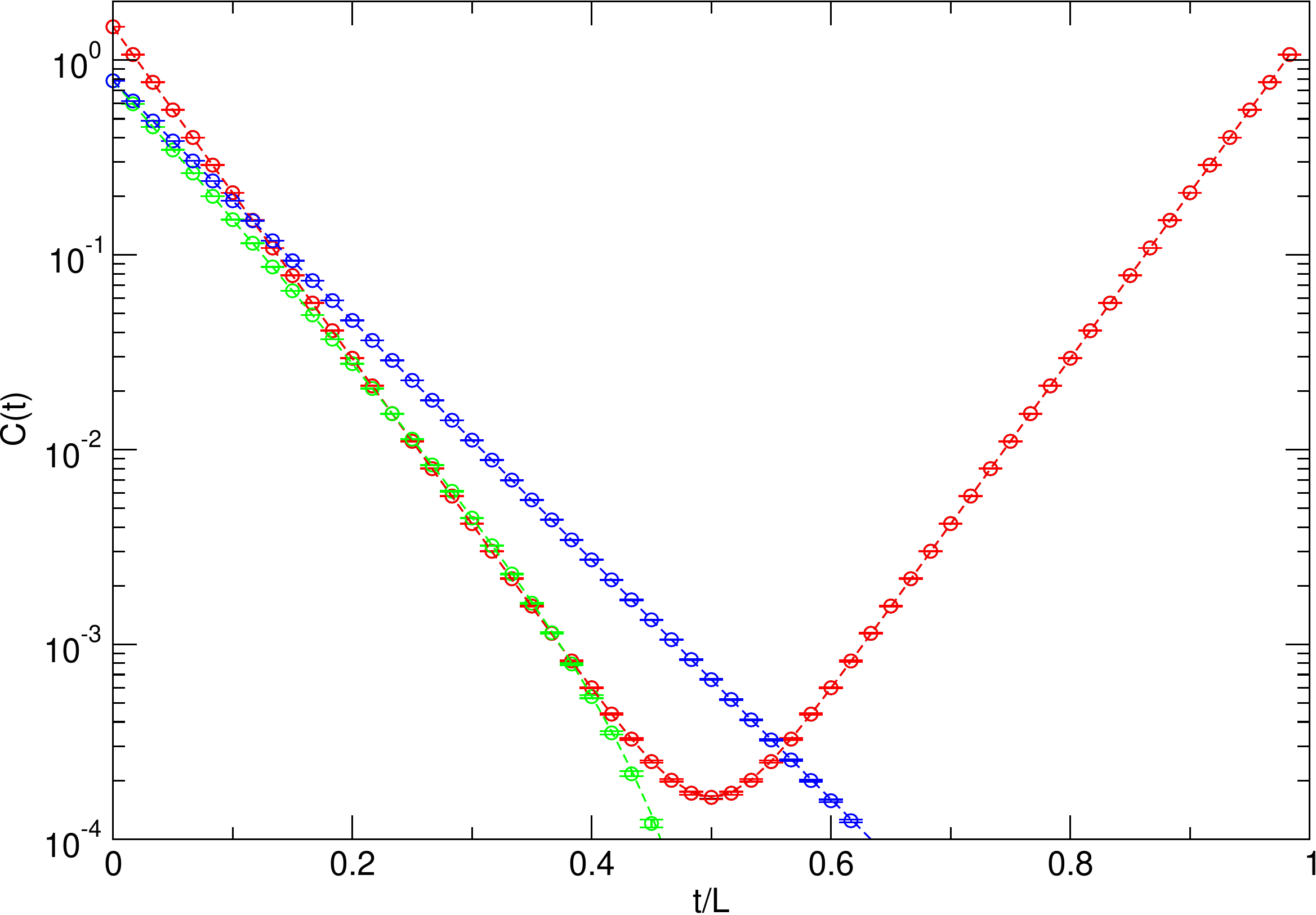}
 \caption{Unbroken supersymmetric quantum mechanics, standard
    discretisation. Bosonic correlation function for antiperiodic
    (black) and periodic b.c.~(red) (lying on top of each other) and
    fermionic correlation function for antiperiodic (green) and
    periodic b.c.~(blue) for $\mu L=10$ at coupling $f_u = 1$. The
    dashed lines are the exact results from \cite{Baumgartner:2015qba}.}
  \label{fig_count_corr_MC}
\end{figure}

First, we consider the standard discretisation with the superpotential
$P_u$ such that supersymmetry is unbroken.  As a first observable, we
show the results for the bosonic and fermionic correlation functions
for $\mu L = 10$, $L/a = 60$ and $f_u = 1$ for $Z_{a} = 10^7$ in
figure \ref{fig_count_corr_MC}.
This is essentially the same plot as figure 10(b) in our second paper
\cite{Baumgartner:2015qba}, but now with the additional data from the
Monte Carlo simulation and plotted on a logarithmic scale. Note that
we use the notation $x=t$ in accordance with
\cite{Baumgartner:2015qba}. The simulation indeed reproduces the exact
result within very small statistical errors which demonstrate the
efficiency of the algorithm. The exponential error reduction is due to
the use of the improved estimators for the two-point function which
are available in the context of the worm algorithms. The improvement
is particularly impressive for the fermionic correlator where the
error reduction allows to follow the correlator over more than seven
orders of magnitude without loss of statistical significance. In fact
the relative error for the lowest value of the fermionic correlator is
still only $4\%$.

\begin{figure}
 \centering
 \includegraphics[width = 0.8\textwidth]{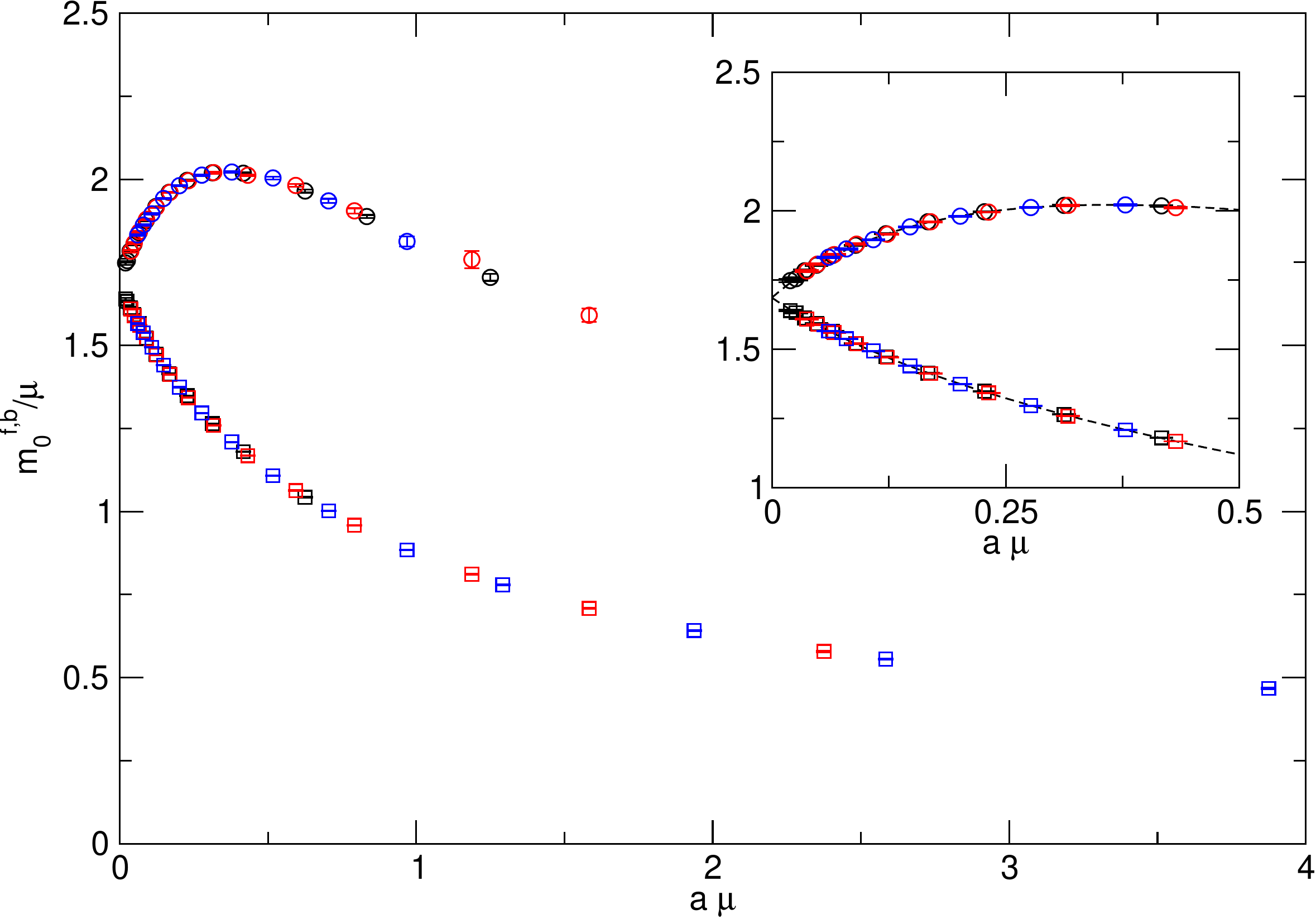}
 \caption{Unbroken supersymmetric quantum mechanics, standard
   discretisation. Continuum limit of the lowest bosonic (circles) and
   fermionic (squares) mass gap for $\mu L = 10$ (black), $\mu L = 19$
   (red), $\mu L = 31$ (blue) and $f_u = 1$. The inset shows a
   detailed comparison with the exact results (dashed lines).}
  \label{fig_count_masses_MC}
\end{figure}

As a second example, we show the mass gaps for different $\mu L$ at a
coupling $f_u = 1$ with statistics of $Z_{a} = 10^6$ in figure
\ref{fig_count_masses_MC}.
The $\mu L$ considered are in the region where thermal effects are
negligible and essentially only $Z_0$ contributes to the total
partition function, such that $Z_{a} \simeq Z_0$.  We extract the
masses from the asymptotic behaviour of the correlation function at
large $t$, i.e., we extract the lowest energy gap. Because of the
extremely good signal-to-noise ratio the asymptotic behaviour can be
truly reached and, in doing so, systematic errors from contributions
of excited states are essentially excluded.  Of course, we know from
our exact results that the overlap of the simple operators we use to
construct the two-point function is close to maximal. This is clearly
visible in figure \ref{fig_count_corr_MC} where we observe an almost
purely exponential decay for all $t/L$.  Because the energy gaps are
independent of $\mu L$, they are expected to fall on top of each other
for all values of $\mu L$ at fixed lattice spacing $a\mu$. This is
indeed the case within our numerical accuracy, and the extracted
masses, when expressed in units of $\mu$, indeed extrapolate to the
correct zero-temperature continuum limit. The inset of figure
\ref{fig_count_masses_MC} shows a detailed comparison of the
simulation results with our exact solution from
\cite{Baumgartner:2015qba} represented by the dashed line and we
observe a beautiful agreement even very close to the continuum.

Next, we consider the action with counterterm and the superpotential
$P_b$ for which the supersymmetry is broken. In this case we encounter
an issue concerning the potential non-positivity of the weights which
we already mentioned in Section \ref{sec:weights arbitrary
  superpotential}. This potentially dangerous sign problem is not of
fermionic origin, but is instead related to the bond formulation of
the bosonic degrees of freedom. As a matter of fact it occurs already
in the purely bosonic system, independent of the dimensionality of the
system. However, negative weights only occur in a region of parameter
space which becomes irrelevant towards the continuum limit. In that
sense, the sign problem is a lattice artefact and can be avoided
straightforwardly.  Nevertheless, in order to eliminate any systematic
error we deal with this bosonic sign problem by incorporating the sign
of the configuration into the observables, even though it has no
practical consequences.

\begin{figure}
 \centering
\includegraphics[width = 0.8\textwidth]{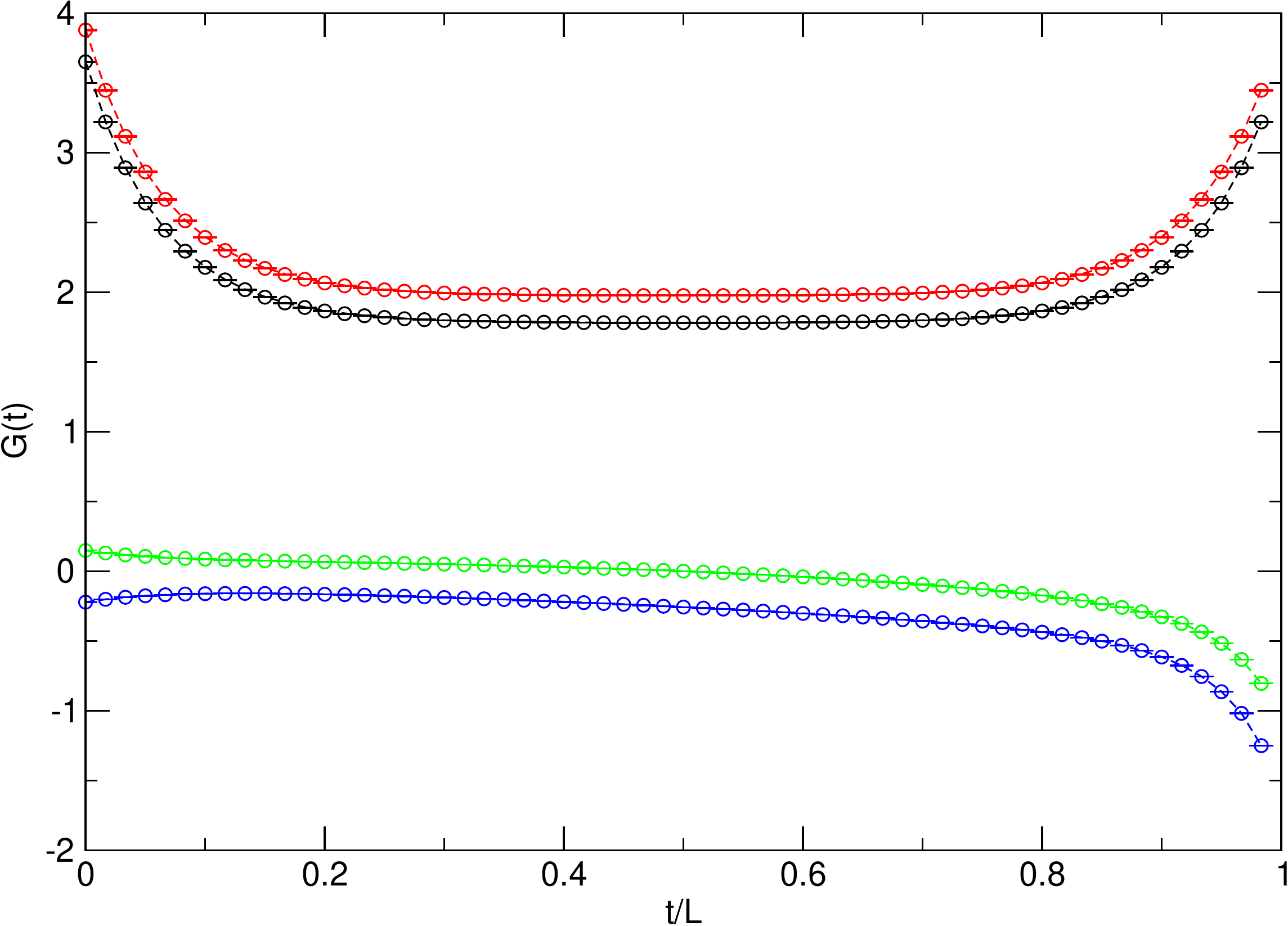}
 \caption{Broken supersymmetric quantum mechanics, standard
   discretisation. The bosonic two-point function for antiperiodic
   (black) and periodic b.c.~(red) and the fermionic one for
   antiperiodic (green) and periodic b.c.~(blue) for $L/a = 60$, $\mu
   L=10$ at coupling $f_b = 1$. The dashed lines are the exact results
   from \cite{Baumgartner:2015qba}.}
  \label{fig_count_G_MC}
\end{figure}

As a first observable in the broken case, we show the bosonic and
fermionic two-point functions, $\langle \phi_t \phi_0 \rangle$ and
$\langle \psi_t \psibar_0 \rangle$, for periodic and antiperiodic
b.c.~for $\mu L = 10$ at fixed coupling $f_b = 1$ in figure
\ref{fig_count_G_MC} for a statistics of $Z_{a} = 10^8$.
The exact results from \cite{Baumgartner:2015qba} are shown as dashed
lines. The simulation yields results which agree with the exact
results within the very small statistical errors on the level of
1$\permil$. Note that the correlators for periodic and antiperiodic
b.c.~are constructed a posteriori from the simulation results in the
bosonic and fermionic sectors $Z_0$ and $Z_1$, respectively, and it is
crucial to sample the relative weight between the two sectors
correctly in order to get the final values right. The relative
sampling is solely in the responsibility of the fermion simulation
algorithm. Our results in figure \ref{fig_count_G_MC} show that the
open fermion string algorithm indeed transits sufficiently well
between the two sectors.

This statement can be made more quantitative by looking at the ratio
$Z_{p}/Z_{a}$ which represents the Witten index in our field theoretic
setup.  From our exact results in \cite{Baumgartner:2015qba} we expect
a nonzero Witten index at finite lattice spacing which however
extrapolates to zero in the continuum limit. So the behaviour of the
algorithm towards the continuum limit is particularly interesting,
because for vanishing lattice spacing the would-be Goldstino at finite
lattice spacing turns into a true, massless Goldstino. In such a
situation one usually encounters critical slowing down of the
simulation algorithms, such that the errors on the results grow large
and the results become unreliable. The massless Goldstino is directly
related to the tunnelling between the bosonic and the fermionic sector,
and the reproduction of a Witten index $W=0$ in the continuum with
small errors is hence a true demonstration of the efficiency of the
open fermion string algorithm to transit between the bosonic and
fermionic sector. In addition, we know from \cite{Baumgartner:2015qba}
that the lattice artefacts are exponentially enhanced towards zero
temperature and it is interesting to see how the simulation algorithm
handles this situation at coarse lattice spacing.

\begin{figure}
 \centering
\includegraphics[width = 0.8\textwidth]{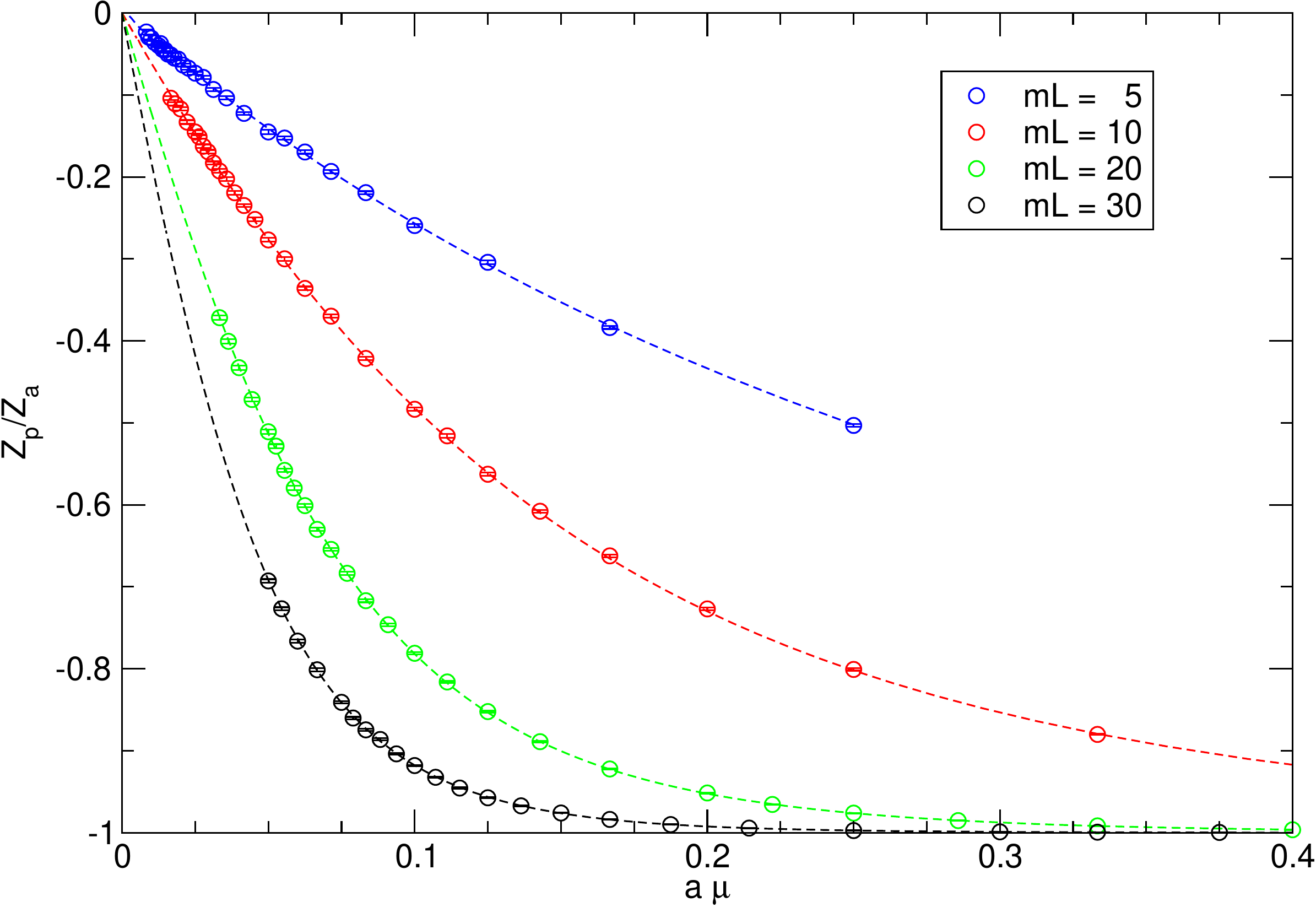}
\caption{Broken supersymmetric quantum mechanics, standard
  discretisation. Continuum limit of the partition function ratio
  $Z_{p}/Z_{a}$, i.e., the Witten index, for $\mu L = 5$ (blue), $\mu
  L = 10$ (red), $\mu L = 20$ (green), $\mu L = 30$ (black) at fixed
  coupling $f_b = 1$. The dashed lines are the exact results from
  \cite{Baumgartner:2015qba}.}
  \label{fig_Zp_Za_count_b_MC}
\end{figure}

In figure \ref{fig_Zp_Za_count_b_MC} we show the ratio $Z_{p}/Z_{a}$
as a function of the lattice spacing $a \mu$ for different values of
$\mu L$ at fixed coupling $f_b = 1$.
For this quantity, too, the simulation yields results which agree with
the exact results within the small statistical errors. Moreover, the
efficiency of the algorithm does not appear to deteriorate towards the
continuum limit or for small values of $\mu L$ where the Witten index
is very close to zero. This can for example be seen from the fact that
the errors obtained with fixed statistics essentially remain constant
towards the continuum limit and are also independent of the system
size.  This nicely demonstrates the efficiency of the algorithm also
for a system with broken supersymmetry.

\begin{figure}
 \centering
\includegraphics[width = 0.8\textwidth]{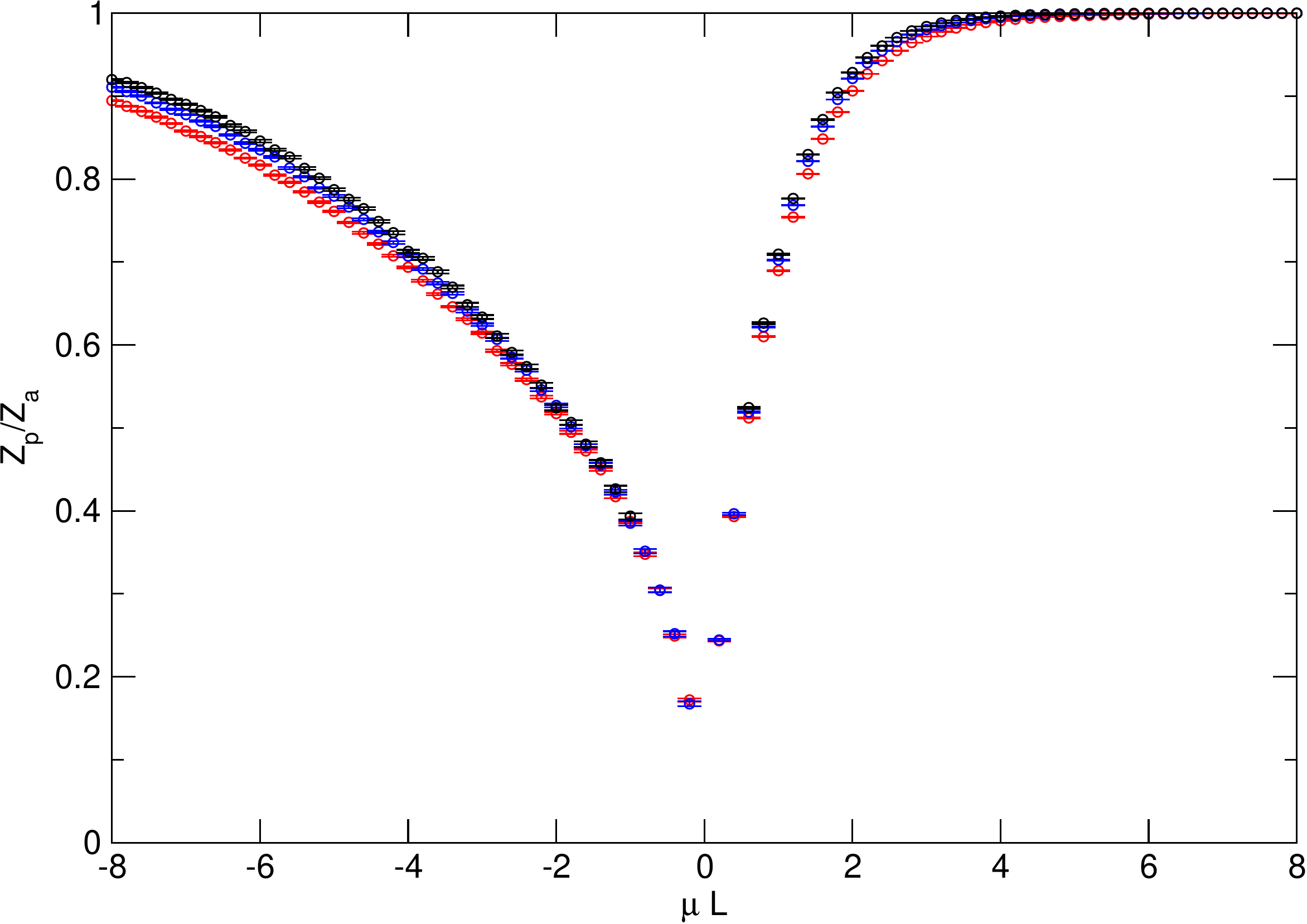}
\caption{Unbroken supersymmetric quantum mechanics, $Q$-exact
  discretisation. $Z_{p}/Z_{a}$ as a function of $\mu L$ for $L/a =
  16$ (red), $L/a = 32$ (blue), $L/a = 64$ (black) at fixed coupling
  $f_u = 1$.}
  \label{fig_Q1_ex_phase_MC}
\end{figure}

The last system we investigate with the worm algorithm is unbroken
supersymmetry formulated with the $Q$-exact action\footnote{For Monte
  Carlo simulations using the $Q$-exact action for broken
  supersymmetry, we encounter the very same problems we ran into in
  the transfer matrix approach. The bond occupation number grows
  extremely large even on small lattices and for coarse lattice
  spacings such that the generation of reliable results turns out to
  be impossible.}.  We first consider the ratio of partition functions
$Z_{p}/Z_{a}$ which in the limit of $\mu L \rightarrow \infty$ yields
the Witten index. From a simulational point of view, the ratio
essentially calculates the fraction of configurations in sector $\Z_0$
versus the ones in $\Z_1$. For unbroken supersymmetry the system is
almost exclusively in the bosonic sector, and hence the ratio is very
close to one except when the size of the system becomes very small,
i.e., in the high temperature limit. Moreover, from our exact results
in \cite{Baumgartner:2015qba} we know that the lattice artefacts in
this quantity are very small and the continuum limit is not very
interesting. For these reasons, we consider in figure
\ref{fig_Q1_ex_phase_MC} the dependence of the ratio $Z_{p}/Z_{a}$ on
$\mu L$ for different values of the lattice spacing $a/L$ with a
statistics of $Z_{a} = 10^8$.

Also for this quantity, we find that the results agree with the exact
result within the very small statistical errors. Again, the open
fermion string algorithm proves to be very efficient even close to
$\mu L \simeq 0$ where the tunnelling from the bosonic to the fermionic
sector and vice versa becomes important and dominates the behaviour of
the system. Thus, even in this somewhat extreme situation of very high
temperature, the algorithm does not show any signs of critical slowing
down despite the fact that there is a quasi-zero mode in the system.

Note that the algorithm is capable of handling negative bare masses
independent of the discretisation used and
fig.\ref{fig_Q1_ex_phase_MC} is simply also an illustration of this
fact.

\begin{figure}
 \centering
 \includegraphics[width = 0.8\textwidth]{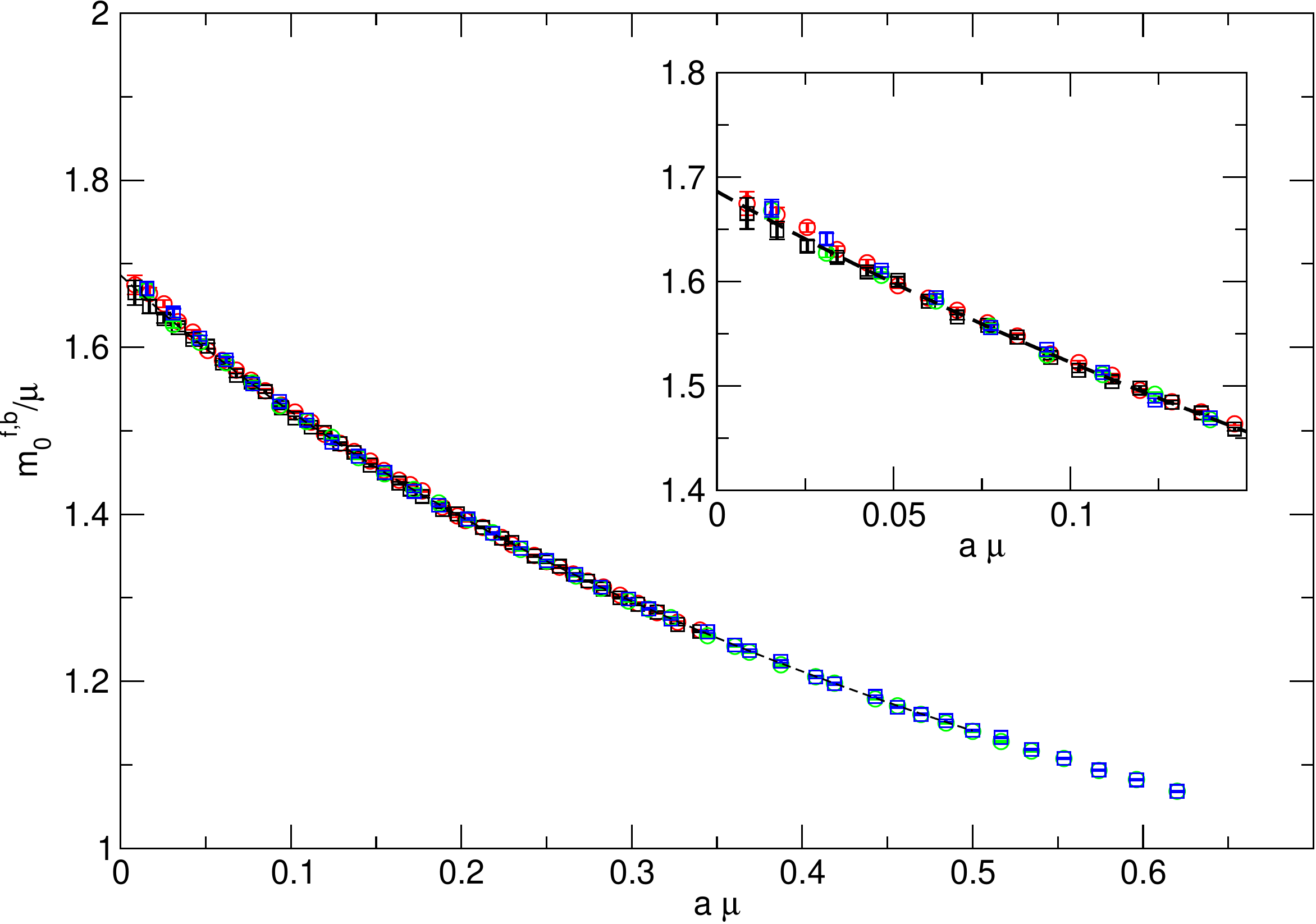}
 \caption{Unbroken supersymmetric quantum mechanics, $Q$-exact
   discretisation. Continuum limit of the lowest bosonic (black
   squares) and fermionic (red circles) mass gaps for $\mu L = 17$,
   and bosonic (blue squares) and fermionic (green circles) mass gaps
   for $\mu L = 31$ at fixed coupling $f_u = 1$. The inset shows a
   detailed comparison with the exact result (dashed line).}
  \label{fig_Q1_ex_masses_MC}
\end{figure}

The last quantity we calculate are the lowest bosonic and fermionic
mass gaps for different $\mu L$ at fixed coupling of $f_u = 1$ from a
statistics of $Z_{a} = 10^6$. The mass gaps are extracted from the
two-point correlation functions exactly in the same way as before for
the standard action, and in figure \ref{fig_Q1_ex_masses_MC} we show
the results of this analysis.
As expected, the masses for the boson and the fermion are indeed
indistinguishable within statistical errors. The degeneracy of the
masses at finite lattice spacing due to the $Q$-exactness of the
action emerges also for the results from Monte Carlo simulations. Note
that the chosen values for $\mu L$ lie well within the region where
thermal effects are negligible and the masses extrapolate nicely to
the correct zero-temperature continuum limit. The inset in figure
\ref{fig_Q1_ex_masses_MC} shows a detailed comparison with our exact
results from \cite{Baumgartner:2015qba} and we again observe beautiful
agreement.

\section{Conclusions}
In this paper we present an algorithm for simulating ${\cal N}=2$
supersymmetric quantum mechanics on the lattice. The algorithm is
based on the reformulation of the system in terms of bosonic and
fermionic bonds, and in essence represents an efficient Monte Carlo
scheme for updating fermionic and bosonic bond configurations. The
updating of the fermionic degrees of freedom is of specific interest,
because this is in general the most challenging part of a
simulation. This is particularly true for systems with broken
supersymmetry, where standard simulation algorithms suffer from
critical slowing down due to the massless Goldstino mode. In addition,
these systems inevitably also suffer from a sign problem related to
the Goldstino and the vanishing Witten index.

In contrast, the fermion simulation algorithm proposed in
\cite{Wenger:2008tq} eliminates critical slowing down by directly
sampling the fermionic two-point correlation function. It is based on
introducing a fluctuating open fermion string which efficiently
updates the bond configurations on all length scales up to the
correlation length associated with the fermionic correlation
function. As a consequence, the fermion string induces frequent
tunnellings between the bosonic and fermionic vacuum when that
correlation length becomes large. Since the two vacua contribute to
the partition function with opposite signs, the frequent tunnelling
guarantees sufficiently small statistical fluctuations for the average
sign, and hence a solution to the fermion sign problem.  In fact, the
more severe the sign problem gets towards the continuum limit, the
more efficiently the algorithm tunnels between the bosonic and
fermionic sectors. This is of course due to the growing correlation
length associated with the vanishing Goldstino mass. The bosonic
degrees of freedom can be expressed in terms of bonds as
well. Therefore, we also give the details of an updating algorithm for
the bosonic bond configurations. Since we consider $Q$-exact
discretisations in addition to the standard one, the algorithm
involves updating generic types of bonds.

The simulation algorithm requires the calculation of the site weights
$Q_F(N)$. Their numerical evaluation, however, turns out to be
numerically unstable for large site occupation numbers $N$. Hence, in
Section \ref{sec:weights}, we devise a computational strategy which
allows to reliably evaluate the ratios of weights for arbitrarily
large occupation numbers. Since this is a generic problem occurring in
the bond formulation of field theories with real scalar fields, such
a computational scheme is useful also in other situations.

Finally, we present a selection of results obtained using the open
fermion string algorithm. We concentrate on two specific realisations
of supersymmetric quantum mechanics, one with broken and one with
unbroken supersymmetry. In addition, we consider both the standard and
the $Q$-exact discretisation. Since exact results are available at
finite lattice spacing from our investigation in
\cite{Baumgartner:2015qba}, we can benchmark our stochastic results
and directly validate them. The calculation of the bosonic and
fermionic correlation functions shows that they can be determined very
accurately over several orders of magnitude. This allows for a very
precise computation of the boson and fermion masses, the latter in
many cases with a smaller error than the former. In general, a
precision of 1\permil \, can be reached with a very modest
computational effort. In systems with broken supersymmetry it is
crucial that the simulation algorithm efficiently samples the relative
weights between the bosonic and fermionic sectors. Our results for the
partition function ratio $Z_p/Z_a$, i.e., the Witten index, show that
this is indeed the case. For fixed statistics, the errors do not grow
towards the continuum limit. In that limit the index gets very close
to zero and the sign problem would therefore be most
severe. Similarly, the error is essentially independent of the system
size, which shows that the sign problem is truly solved.

\bibliography{susyQM_Simulations}
\bibliographystyle{JHEP}
\end{document}